# Micromagnetic modelling of nanorods array-based L1$_0$-FeNi/SmCo$_5$ exchange-coupled composites

V.L. Kurichenko[1], D.Yu. Karpenkov[1] and P.A. Gostischev[2]

[1] National University of Science and Technology 'MISIS', Leninskiy prospect, 4, Moscow, 119991, Russian Federation
[2] L.A.S.E.-Laboratory for Advanced Solar Energy, National University of Science and Technology 'MISiS', Leninskiy prospect 6, Moscow, 119049, Russian Federation

E-mail: vkurichenko@misis.ru

## Abstract

Exchange-coupled nanocomposites are considered as the most promising materials for production of high-energy performance permanent magnets, which can exceed neodymium ones in terms of energy product. In this work, micromagnetic simulations of L1$_0$-FeNi/SmCo$_5$ composites based on the initially anisotropic structure of nanorods array were performed. Texturing effect on magnetic properties was investigated. It was revealed that even 30 % of anisotropy axes misalignment of grains in L1$_0$-FeNi phase would lead to only ≈ 10 % drop of coercivity. To maximize magnetic properties of the composites, parameters of microstructure were optimized for 120 × 120 array of interacting nanorods and were found to be 40 nm nanorod diameter and 12–20 nm interrod distance. The estimated diameter of nanorods and the packing density of the array provide energy product values of 149 kJ m$^{-3}$. Influence of interrod distance on energy product values was explored. Approaches for production of exchange-coupled composites based on anisotropic nanostructures were proposed.

## 1. Introduction

Production of exchange-coupled magnetic materials is considered as one of the most promising methods for enhancement of energy product '|$BH$|$_{MAX}$' in permanent magnets [1]. For exchange-coupled nanocomposites, spins of a soft phase are strongly coupled with spins of a hard phase, which leads to enhancement of coercivity in such material [2]. It is predicted that such composites can even exceed Nd$_2$Fe$_{14}$B-based magnets in terms of energy product, which is a figure of merit in hard magnetic material [3]. Since the discovery of such composites in 1919 by Kneller and Hawig [4] numerous theoretical and experimental research works have been carried out in order to understand the exchange-coupling effect.

There were various articles concerning theoretical aspects of exchange bias [5–7], application of exhange-coupled nanocomposites [8, 9], exchange-coupling features in different types of alloys [10] and related effects in nanostructures [11].

Kneller and Hadwig [4] in their work concluded that the optimal microstructure of exchange-coupled nanocomposites should be a homogeneous distribution of a hard phase in a magnetically soft matrix. Accordingly, dimensions of both phases should be equal with the assumption that the hard phase precipitates are spherical and spatially distributed according to face-centered cubic (fcc) lattice.

Later, other reports predicted the effect of soft-hard phases microstructures on exchange-coupling nanocomposite magnetic behaviour.

For instance, Skomski and Coey analytically revealed that a single ideally aligned soft inclusion in a hard matrix could enhance the remanence magnetization '$M_R$'. For further increase, the microstructure of multiple inclusions should be realized. However, optimal microstructure would be sufficiently small soft regions (in order to avoid low-field nucleation) and a crystallographically oriented hard phase (so it acts as a skeleton for stiffening the magnetization direction of the soft region). For that, the microstructure was suggested to be a disordered two-phase magnet with common $c$-axis through hard regions and multilayered alternating soft-phase layers [12].

Skomski et al. [13] in their further work additionally analyzed such microstructures like a cylinder-shaped soft inclusion, an embedded soft layer in a hard matrix and a free soft layer at the surface. The embedded spherical layer had the highest coercivity compared to other geometries. For example, the spherical case was higher than the layered case by a factor of 4 and 16 for the embedded layers and the free surface layer, respectively. The embedded soft cylinders case was intermediate between the spheres and the layers. The main conclusion was that soft-in-hard geometry is more favourable, which is opposite to Knelled and Hadwig results.

Jiang and Bader [14] suggested that a soft inclusion in a hard matrix microstructure may not be an optimal one for exchange-coupled nanocomposite, as it has the lowest soft/hard phase volume ratio and close-packing efficiency for 3D case (≈ 74 %). Therefore, they assessed different structures for maximizing of $|BH|_{MAX}$, including core/shell structure. It was concluded that multilayered geometry has the highest packing efficiency and spherical soft core/hard shell geometry has the highest nucleation field. However, maximum value of $|BH|_{MAX}$ for a cylindrical soft phase in a hard matrix is close to the corresponding value for the multilayered structure. Additionally, the soft phase linear dimensions for the columnar microstructure could be higher compared to the multilayered case. Therefore, it was suggested that the cylindrical soft/hard microstructure could be an optimal one for exchange-coupled nanocomposites.

In addition, the properties of constituent phases of composites are also very important and must be considered. It was previously shown that only a few nanometers (≈ 5 nm) thick layer of a soft magnetic phase could be effectively coupled with a hard layer [4]. Such small dimensions of constituent composites phases will be hard to manipulate. An analytical solution for the critical soft phase thickness can be presented as an equation [4]:

$$t_{\text{S-Critical}} = \frac{J_{\text{ex}}/M_{\text{S}}}{\frac{2K_H t_H - J_{\text{ex}}}{M_H t_H} - \frac{2K_S}{M_S}} \quad (1)$$

where $J_{ex}$ is the exchange strength, $K_S$ is anisotropy constant of the soft phase, $K_H$ is anisotropy constant of the hard phase, $M_S$ is saturation magnetization of the soft phase, $M_H$ is saturation magnetization of the hard phase, $t_H$ is the hard phase thickness.

This equation reveals that such parameters of the soft phase like saturation magnetization and anisotropy constant can change the critical soft phase diameter. Thus, by using a 'less soft' phase, i.e. semi-hard phase, a critical dimension of the soft phase can be increased [4]. That can solve the problem of small soft phase dimensions and make it less problematic to work with soft-hard exchange coupled nanocomposites.

Ordered FeNi phase (tetrataenite) with L1$_0$ structure can be used as a semi-hard phase in the exchange-coupled nanocomposites. Among all rare-earth free phases, tetrataenite is one of the most promising substitutional magnetic materials. Since its discovery in 1964 [15], various reports have been published concerning tetrataenite production in the laboratory. The main challenge for laboratory synthesis of tetrataenite is attributed to slow diffusion at the chemical temperature of its order-disorder transition, which equals to 573 K. This is the reason why in nature tetrataenite was observed only in meteorites, as they had million years for the transition of a disordered structure to the ordered one [16]. Therefore, methods for laboratory synthesis of tetrataenite mainly focus on increasing the diffusion rate in the low-temperature region. There were different methods reported for tetrataenite production, e.g. neutron bombardment [15], severe plastic deformation by high-pressure torsion [17], cyclic oxi-reduction treatments [18, 19], sputtering [20, 21] annealing of amorphous films [22], high-energy ball milling [23] etc. However, tetrataenite was obtained either in small fractions or not in bulk samples. Recently, tetrataenite production by means of nitrogen insertion and topotactical extraction technique [24] was reported in initial FeNi A1 nanopowders. Single-phased samples had a high order parameter (≈ 0.71) and an anisotropy constant (up to 3 MJ m$^{-3}$). Therefore, this method can be used for formation of tetrataenite in nanocomposites.

In this paper by means of micromagnetic simulations we investigated magnetic properties of FeNi-L1$_0$/SmCo$_5$ exchange-coupled composites based on nanorods array. It is expected that such nanocomposites will show enhanced values of energy product.

## 2. Model description

Calculations of the exchange-coupled nanocomposites were performed in MuMax3, which is a GPU-accelerated software that uses finite-difference discretization for micromagnetic simulations [25].

**Table 1.** Magnetic properties of constituent phases. '$M_S$' is saturation magnetization, '$K_u$' is uniaxial anisotropy constant, '$A_{ex}$' is exchange stiffness parameter.

| Phase | $M_S$, kA m$^{-1}$ | $K_u$, MJ m$^{-3}$ | $A_{ex}$, pJ m$^{-1}$ |
|---|---|---|---|
| FeNi L1$_0$ | 1280 [26] | 0.32 [26] | 11.2 [27] |
| SmCo$_5$ | 860 [28] | 17 [28] | 12.0 [28] |

In our work, we modelled two cases: ε = 10 % and 20 %, which provided the exchange coupling coefficient values of 0.9 and 3.5 mJ m$^{-2}$, respectively. These values lie in the range which is typical for exchange-coupled composites modelled in the literature (from several mJ m$^{-2}$ up to 10 mJ m$^{-2}$) [34]. It has to be noted that the exchange-coupling coefficient can indeed be increased, e.g. by means of introduction of the additional interface layer which enhances the exchange between soft and hard phases. This phenomenon was observed in SmCo/Fe system both experimentally [35] and theoretically [36].

Another essential issue to consider is the hard phase thickness '$t_H$'. Since the exchange-coupling have close-range behaviour and only a small fraction of a hard phase is participating in the effect [32], there is no need to use very thick layers. We proposed to set the upper limit of the hard phase thickness equal to the exchange length of SmCo$_5$, calculated by equation (3), which is 5 nm. Therefore, since MuMax3 performs best with power-of-two sizes [25], 4 nm thickness of SmCo$_5$ was used.

The soft phase nanorods diameter is limited by production technique. We suggest to produce nanorods by means of electrodeposition technique on polycarbonate membranes, which act as templates [37]. Membranes with variety of pore sizes are available on the market, therefore we will model different morphology of nanorods (20, 40, 60, 80 and 100 nm) to investigate its influence on magnetic properties of the exchange-coupled nanocomposites. Nanorods length was set to 1 μm. The cell size used in calculations was 2 nm.

## 3 Results and discussion

### 3.1 Estimation of optimal nanorods diameter

At the first stage, we performed modelling of a isolated nanorod in order to estimate its optimal diameter '$d_N$' for yielding maximized magnetic properties. The structure was grained and textured. Grains were introduced by means of Voronoi tesselation. The grain size was set to 10 nm. Exchange coupling between grains was reduced by 10 % to simulate the grain boundary impact. Random 10 % variation of anisotropy constant values were set in each grain of soft and hard phases. Magnetization vector and anisotropy axes of soft and hard layers grains were parallel to z-axis and both corresponded to <001> direction. In all calculations magnetic field was applied parallel to nanorods axis.

Figure 1 shows coercive field '$H_C$' and remanent magnetization $M_R$ versus nanorods diameter dependence for initial noncovered nanorods (red) and SmCo$_5$-covered composites with 10 % (green) and 20 % (blue) exchange-coupled spins, respectively. Obtained results indicate that the optimal diameter of the isolated FeNi-L1$_0$/SmCo$_5$ nanorod with given properties (see table 1) is 40 nm. Using nanorods with 40 nm in diameter will not lead to the significant decrease of $M_R$, as it is observed for nanorods with $d_N$ = 20 nm. At the same time, nanorods with $d_N$ = 40 nm will still have enchanced $H_C$ values compared to nanorods with higher $d_N$. Decrease of $M_R$ with nanorod diameter variation is attributed with the hard magnetic layer introduction. The latter leads to diluting of L1$_0$-phase which has high saturation magnetization with hard magnetic SmCo$_5$ phase possessed lower values of $M_S$. Higher $M_R$ values for covered nanorods with ε = 20 % in comparison with the values for ε = 10 % are explained by the fact that in the former exchange-coupling is stronger, leading to suppression of magnetization reversal process in the soft magnetic phase. Values of $H_C$ obtained for 40 nm thick nanorods are 282, 392 and 560 kA m$^{-1}$ for initial noncovered nanorods and SmCo$_5$-covered nanorods with ε = 10 % and ε = 20 %, respectively. Corresponding values of $M_R$ are 1279, 1149 and 1172 kA m$^{-1}$.

### 3.2 Texturing effect on magnetic properties

In the previous section, the micromagnetic calculations were done for a textured soft magnetic phase in an isolated single nanorod-based composite. We modelled structure with anisotropy axes of all grains aligned in <001> direction.

However, it is well known that texturing can be a quite challenging task. Therefore, in this section, we considered more close-to-experiment structure, i.e. grained structure with different anisotropy axis distribution in the soft magnetic phase. In particular, we modelled isolated FeNi-L1$_0$/SmCo$_5$ nanorods, whose grains in the tetrataenite phase have anisotropy axes distributed with a maximum deviation from <001> direction of 30, 60 and 80 degrees. In the following text those distributions will be referred according to their maximum deviation value '$α_{dev}$'. The hard magnetic SmCo$_5$ phase was kept textured along <001> direction. Series of calculations were conducted in each case for obtaining averaged values. Calculation results for noncovered isolated rod are presented in figure 2.

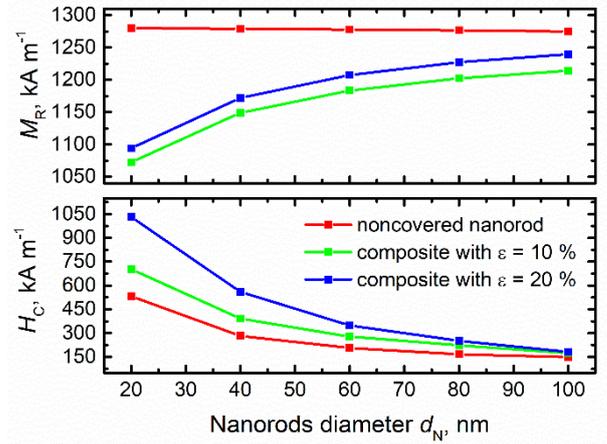

**Figure 1.** Coercivity and remanent magnetization dependences of the exchange-coupled composites on the diameter of nanorods.

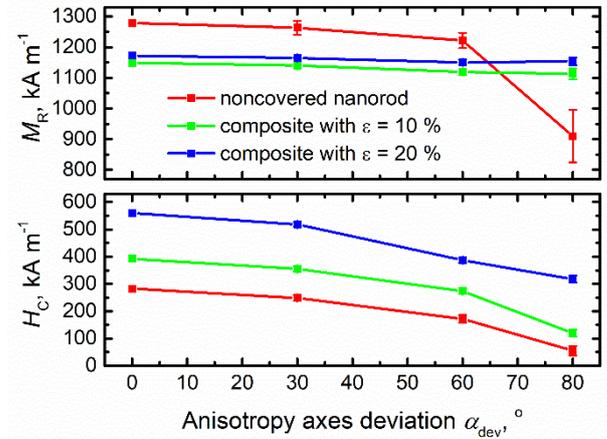

**Figure 2.** Coercivity and remanent magnetization dependences of the exchange-coupled composites on anisotropy axes deviation in the soft magnetic phase.

Results reveal a drop in $M_R$ values, especially for $α_{dev}$ = 80°. Such behaviour was expected, as this value of deviation can be attributed to the structure with almost random anisotropy axes distribution in grains. However, for the exchange-coupled nanorod composites, such decrease of $M_R$ is not so dramatic, as spins of the soft magnetic phase are coupled to the spins of the hard magnetic phase, resulting in the soft phase magnetization vectors alignment in accordance to textured grains of the hard magnetic layer.

Observed reduction of $M_R$ with $α_{dev}$ increase can be explained by competition between shape and magnetocrystalline anisotropies. The shape anisotropy constant $K_d$ can be calculated as follows [38]:

$$K_d = \frac{1}{2}\frac{NM_s^2}{\mu_0} \quad (4)$$

where $\mu_0$ is the magnetic constant, $N$ = ½ is the radial demagnetizing factor of a long cylinder (nanorod).

The estimated value of $K_d$ for for tetrataenite phase is 0.51 MJ m$^{-3}$. Magnetocrystalline anisotropy constant of FeNi-L1$_0$ phase $K_u$ equals to 0.32 MJ m$^{-3}$.

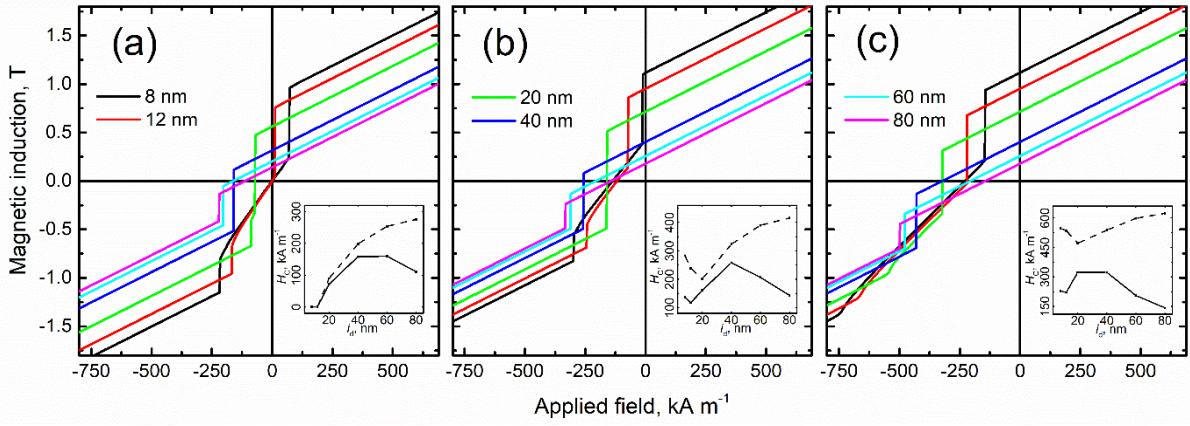

**Figure 3.** Calculated $B(H)$ demagnetizing curves for 40 nm in diameter 120 × 120 nanorods array with different interrod distance. a) FeNi $L1_0$ nanorods array. b) FeNi $L1_0$/SmCo$_5$ composites with 10 % exchange-coupled spins. c) FeNi $L1_0$/SmCo$_5$ composites with 20 % exchange-coupled spins. Insets on all figures show coercive field vs interrod distance dependence (solid lines correspond to $H_C^B$, dashed lines – $H_C^M$).

It is seen that both values of the competing anisotropy constants are in the same order of magnitude, i.e. $K_d \approx K_u$. Therefore, when strong magnetocrystalline anisotropy is introduced, additional directions corresponding to the easy axes of each individual grains form the set of randomly distributed local minimums.

The latter leads to the appearance of intermediate energetically favourable directions for magnetization vector and, consequently, to decreasing of coercivity and remanent magnetization in the FeNi $L1_0$-phase nanorod.

It is worth noting that deviation degree of $\alpha_{dev} = 30°$ for the exchange-coupled composites with $\varepsilon = 10$ % and 20 % leads to decrease of $H_C$ values only of 9.5 and 7.5 %, respectively, in comparison with textured structure, i.e. structure with $\alpha_{dev} = 0°$. Therefore, the exchange-coupling effect allows usage of not perfectly textured soft phase without a significant drop of coercivity, thus providing higher energy product values.

### 3.3 Energy product of nanorods array composites

In this section, we modelled magnetic properties of FeNi-$L1_0$/SmCo$_5$ echange-coupled composite based on nanorods array. The array consisted of 120 × 120 tetrataenite nanorods, covered with a thin layer (4 nm) of textured ($\alpha_{dev} = 0°$) SmCo$_5$ phase. FeNi $L1_0$ phase anisotropy axes distribution in grains was also set to 0° in for calculation of the maximum possible energy product for the investigated composites. The distance between nanorods (interrod distance) '$i_d$' is another important parameter to consider in order to maximize magnetic properties of the exchange-coupled nanocomposites. Lower $i_d$ value will have an effect on magnetic dipolar interactions between nanorods [39]. Moreover, since demagnetizing field is higher around nanorod tips, nucleation will happen at lower fields and will favour reversal in the adjacent nanorods. This will result in reduced loop coercivity and squareness [40, 41]. On the other hand, change in $i_d$ will also have an effect on magnetization values, because a decrease of packing density leads to dilution of magnetic material with non-magnetic spacer and subsequent reduction of volume magnetization. Hence, interrod distance plays a key role in performance of the exchange-coupled anisotropic composite-based materials in terms of energy product values. Therefore, it is essential to estimate the optimal distance between nanorods for maximizing $|BH|_{max}$ value of nanorods array. In our calculations $i_d$ varied in the range from 8 to 80 nm. Figure 3 demonstrates $B(H)$ demagnetizing curves for noncovered nanorods and exchange-coupled composites with various $i_d$ distances.

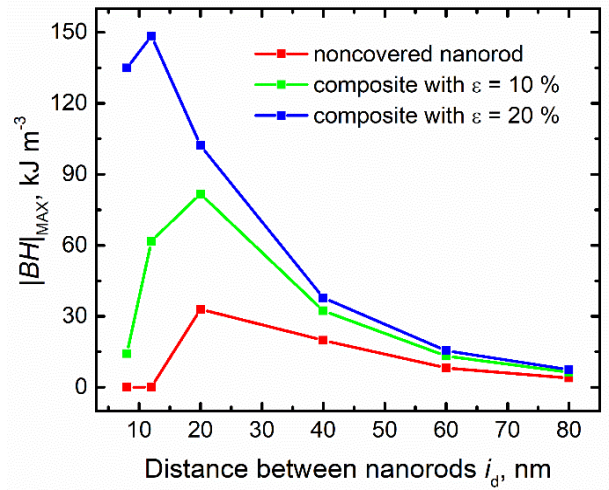

**Figure 4.** Energy product dependence on different interrod distance for FeNi $L1_0$ phase-based nanorods arrays.

Calculations reveal that increase of the interrod distance is accompanied by a rise of absolute value of the nucleation field '$H_N$', which is explained by reduced magnetostatic interaction of the nanorods. In turn, increased magnetostatic interaction leads to zero $H_C$ value for noncovered tetrataenite nanorods arrays in the range of $i_d$ below 12 nm.

It is known that there are at least two types of $H_C$ values [42]. The first one is derived out from $M(H)$ curves, and the second one can be obtained from $B(H)$ curves. These values are referred as '$H_C^M$' and '$H_C^B$', respectively. Insets on figure 3 show the interrod distance influence on $H_C^M$ (dashed lines) and $H_C^B$ (solid lines) values. It is seen that reduced magnetostatic interaction between nanorods via increased interrod distance leads to enhancement of $H_C^M$ values up to the corresponding value for the isolated nanorod. However, for energy product calculations $H_C^B$ values are important and they do not follow the same behaviour as $H_C^M$ values.

Demagnetizing curves on figure 3 show that, e.g. for $i_d = 60$ nm (cyan) critical switching filed values (intrinsic coercivity), that correspond to rapid drops on $B(H)$ curves, increased for the exchange-coupled nanorods arrays. However, since those points are moved to the third quadrant, they are not relevant for energy product estimation.

A pronounced maximum of $H_C^B$ for both covered and noncovered composites in the vicinity of $i_d = 40$ nm is seen in figure 3 insets. However, the maximum of remanence is observed in the interrod distance range of $i_d < 40$ nm. Hence, there should be an optimal $i_d$ value where $H_C^B$ and magnetization are still comparatively high in order to yield maximized energy product.

Figure 4 shows $|BH|_{max}$ vs $i_d$ dependence for tetrataenite nanorods array and exchange-coupled composites with different percentage of exchange-coupled spins. Analysis of

It is worth noting that the calculations revealed tendency of the optimal interrod distance reduction with the increasing percentage of exchange-coupled spins '$\varepsilon$'. In turn, in the case of materials without any nonmagnetic spacing, contribution of the shape anisotropy becomes negligible. Thus, two possible ways of performance enhancement of the exchange-coupled composites can be proposed: i) one should provide the highest possible values of $\varepsilon$ and consequently increase the interface exchange coupling coefficient via engineering of the soft/hard phases interface. The latter can be done, for instance, by means of avoiding of the surface oxidation or by introduction of intermediate layers, which can reduce lattice mismatch or increase exchange-coupling parameter directly [35, 36]. On the other hand, as it was shown, to maximize the energy product of such strong-coupled composites, the packing density should be close to 100%. Hence, in this approach there are no reasons to produce anisotropic composites because the shape anisotropy contribution became zero due to enhanced magnetostatic interaction between nanorods in 'full-dense' material; ii) because of imperfection of composites production conditions and complexity of controlling the percentage of exchange-coupled spins, one should provide nonmagnetic spacer between nanorods to fully utilize the shape anisotropy of anisotropic composites. The latter, as was mentioned above, will lead to decreasing of volume magnetization due to diluting of the magnetic phases with the nonmagnetic spacer. In this case, the optimal interrod distance should be adjusted in order to maximize energy product.

## Conclusions

Micromagnetic modelling of exchange-coupled FeNi L1$_0$/SmCo$_5$ nanorods array-based nanocomposites was performed. Nanorods diameter of 40 nm was found to be optimal. Calculations revealed that 30 % misalignment in anisotropy axes of the soft magnetic phase grains will not significantly decrease coercivity values ($\approx$ 10 % decrease was observed) due to the exchange-coupling effect. Optimal interrod distance, calculated for $120 \times 120$ array of tetrataenite nanorods with 40 nm diameter and 1 μm length, covered with 4 nm thick SmCo$_5$-layer, is equal to 12–20 nm. Estimated energy product values for exchange-coupled composite with 10 % and 20 % exchange-coupled spins are 82 and 149 kJ m$^{-3}$, respectively. According to obtained results, we can propose two approaches for production of the exchange-coupled composites with enhanced magnetic properties: i) provide the highest possible percentage of exchange coupled spins. In this case the optimal interrod distance should be close to zero and contribution of the shape anisotropy will become neglible; ii) use anisotropic nanostructured composites with appropriate nonmagnetic interrod spacer thickness in order to utilize the shape anisotropy of nanorods, while maintaining high volume magnetization values.

## Acknowledgements

This work was supported by the Russian Science Foundation grant No. 18-72-10161. D.Yu.K. gratefully acknowledges the financial support of the Ministry of Science and Higher Education of the Russian Federation in the framework of Increase Competitiveness Program of MISiS (Grant No. P02-2017-2-6).